\documentclass[traditabstract,twocolumn]{aa}
\usepackage{natbib}
\usepackage{graphicx}
\usepackage{txfonts}

\newcommand{\filter}[1]{\mbox{\it #1\/}}              

\begin{document}
\title{The low--mass Initial Mass Function in the Orion Nebula cluster based on  HST/NICMOS III imaging}


\author{M. Andersen 
\inst{1}
\and
M.~R. Meyer \inst{2}
\and 
M. Robberto \inst{3}
\and 
L.~.E. Bergeron \inst{3} 
\and 
N. Reid \inst{3}}

\institute{Research \& Scientific Support Department, ESA ESTEC, Keplerlaan 1, 2200 AG Noordwijk, The Netherlands  
\email{manderse@rssd.esa.int}
\and 
Institute for Astronomy ETH, Physics Department, HIT J 22.4, CH-8093 Zurich, Switzerland
\and
Space Telescope Science Institute, Baltimore, MD 21218, USA 
}




   \date{}

\begin{abstract}
{We present deep HST/NICMOS Camera 3 \filter{F110W} and \filter{F160W} imaging of a 26\arcmin$\times$33\arcmin, corresponding to 3.1pc$\times$3.8pc,  non-contiguous field towards the Orion Nebula Cluster (ONC). 
The main aim is to determine the ratio of low--mass stars to brown dwarfs for the cluster  as a function of radius out to a radial distance of  1.5pc. 
The sensitivity of the data outside the nebulous central region is \filter{F160W}$ \le21.0$ mag, significantly deeper than previous studies of the region over a comparable area. 
We create an extinction limited sample and determine the ratio of low--mass stars (0.08--1 M$_\odot$) to brown dwarfs (0.02--0.08M$_\odot$ and 0.03--0.08M$_\odot$) for the cluster as a whole and for several annuli. 
The ratio found for the cluster within a radius of 1.5pc is $R_{02}=N(0.08-1M_\odot)/N(0.02-0.08M_\odot)=1.7\pm0.2$, and  $R_{03}=N(0.08-1M_\odot)/N(0.03-0.08M_\odot)=2.4\pm0.2$, after correcting for field stars.  
The  ratio for the central 0.3pc$\times$0.3pc region down to 0.03M$_\odot$  was previously found to be $R_{03}=3.3^{+0.8}_{-0.7}$, suggesting the low--mass content of the cluster is mass segregated. 
We discuss the implications of a gradient in   the ratio of stars to brown dwarfs in the ONC in the context of  previous measurements of the cluster and for other nearby star forming regions. 
We further discuss the current evidence for variations in the low--mass IMF and primordial mass segregation. }
\end{abstract}


\maketitle



\section{Introduction}
A central question in astrophysics is whether the Initial Mass Function (IMF) is universal or if it is a function of star formation environment. 
The answer is crucial for our understanding of the buildup of galaxies  and provides clues to how stars and brown dwarfs are formed. 
 The stellar part of the IMF appears to be universal based on the comparison of the field IMF \citep{chabrierreview} to clusters within 2 kpc \citep[e.g.][]{meyer00}. 
\citet{bastian10} reached the same conclusion with the possible exception of the  Taurus Dark Could \citep{luhman09}. 
For more massive Galactic clusters and for low--metallicity clusters in the Large Magellanic Clouds the IMF is well represented by  a power--law to the sensitivity and crowding limit of $\sim$1M$_\odot$ \citep[e.g.][]{stolte,gouliermis,brandner,andersen09}.

The situation is less clear for the brown dwarf  part of the  IMF\@. 
\citet{briceno} and \citet{luhman} suggested a higher ratio of stars to brown dwarfs in Taurus  than the Orion Nebula Cluster (ONC). 
However, \citet{guieu} utilized a larger area survey of Taurus and found the ratio of stars to brown dwarfs to be similar for the two regions \citep[c.f.][]{luhman09}. 
\citet{andersen08} combined the measured IMF's for six nearby star--forming regions and the Pleiades and showed that there is  no strong evidence for variations in the derived ratio of low--mass stars to brown dwarfs  for the regions 
and that the combined ratio  for the 7 regions showed that the underlying IMF was falling  into the brown dwarf regime. 
 Conversely, a handful of  studies  for OB associations over large areas of the sky have indicated an IMF that does not show evidence for a falling mass function in the brown dwarf regime \citep[e.g.][]{lodieu}. 
The reason for the apparent discrepancy is not clear but could be due to either differences in the star formation process or due to dynamical evolution of  the clusters. 
For several of the clusters compiled by \citet{andersen08} only the central regions were studied (e.g. a 0.6pc square box  for the ONC and  0.24pc square box for Mon R2) and the  measured IMF in the centre can  thus be biased if mass segregation is present. 
Large field of view surveys of young star forming regions are needed to directly compare with the OB associations. 

One nearby region of particular interest is the ONC.
So far, the cluster content has only systematically been studied deep into the brown dwarf regime for the central $\sim$5\arcmin\ square \citep{HC00,muench,slesnick}, corresponding to $\sim$ 0.6pc square for a distance of 414pc \citep{menten}. 
 Larger area, less deep, studies of the stellar population have indicated at least the higher mass objects in the  ONC is  mass segregated \citep{HH}. 
Thus, the low--mass IMF derived for the central area can be  biased towards higher mass objects and may not be representative of the cluster as a whole. 
However, no studies have combined deep observations in the near--infrared together with a large field of view coverage such that the brown dwarf content can be studied in detail as a function of radius and to determine the global ratio of stars to brown dwarfs. 
Here we present HST NICMOS Camera 3 \filter{F110W} and \filter{F160W} (similar to the ground--based \filter{J} and \filter{H} bands) non--contiguous observations of the ONC area. 
The observations cover the cluster from the very centre out to a radius of 2.55pc  and were obtained as part of the  HST Orion Treasury Project (GO  10246, PI Robberto).

The paper is structured as follows. 
 Section 2  presents the observations and  data reduction. 
The source detection and the photometry is discussed in Section 3. 
In Section 4 we derive the ratio of low mass stars to brown dwarfs  as a function of radius in the cluster  and we discuss the evidence for the lower mass cluster content to be mass segregated. 
In Section 5 we compare the findings for the ONC with the similar results from other nearby star forming regions. 
Finally, we conclude in Section 6. 
\section{Observations} 
Regions of the ONC were observed using the HST/NICMOS Camera 3 during the  HST Treasury Program covering the Orion Nebula Cluster (GO10246). 
The non--contiguous nature of the NICMOS Camera 3 observations stems from the fact they were observed in parallel with the ACS and WFPC2 Treasury Project observations. 
The field--of--view of the NICMOS Camera 3 is 51\farcs2$\times$51\farcs2\ and the pixel scale is 0\farcs2. 
Since historically, the NICMOS Camera 3 is slightly out of focus, there is 
a slight degradation of the point spread function (PSF) which in turn improves the sampling of the PSF. 

A total of 102 regions were covered in both the \filter{F110W} and \filter{F160W} bands. 
The regions were typically build up as a 4 piece mosaic in the \filter{F160W} band and a 5 piece mosaic in the \filter{F110W} band utilizing the Field Offset Mechanism (FOM, see below). 
In one point, a single 51\farcs2$\times$51\farcs2 image was observed in both filters. 
The integration time of each image was 192 seconds in \filter{F160W} and 256 seconds in  \filter{F110W}, resulting in a total integration time per pixel that varies between 192--762 seconds for the  \filter{F160W} observations and 256--1280 seconds for the \filter{F110W} observations, with the central parts  of each sub--mosaic being imaged for the full amount of time. 
Around 55\%  of the mosaics are covered for 192 seconds in the \filter{F160W} band, and 40\% for  380 seconds. 
For the \filter{F110W} filter, around 45\% is covered for 256 seconds, and 40\% for 512 seconds. 

The large dithering pattern was possible due to the FOM, which however introduces some vignetting. 
Fortuitously, a 3$\times$3 NICMOS Camera 3 mosaic of the Trapezium was obtained in 1998 \citep{luhman_ONC}. 
We were able to make ratios of one 
vignetted FOM-dithered exposure to that unvignetted exposure. 
A smooth
function was fit to the vignetted edge in the ratio image and then expanded into a
2-D vignetting correction image (one per filter for \filter{F110W} and \filter{F160W}) that was applied to all affected exposures.
The raw images have further  been processed with the appropriate dark and flat frames as described on the STScIs web pages. They were further corrected for bad pixels determined from sigma clipping of the median stacked image. 

Offsets were determined between the individual images of  each region by identifying stars in common and determining the centroid. 
 The frames were then combined using the drizzle software that takes the distortion of the NICMOS Camera 3 frames into account and combines the frames with sub--pixel accuracy.  
All the source detection, astrometry,  and photometry is being done on the final mosaic for each  region.

\section{Source detection and photometry}
We describe the source detection and photometry process. 
The photometry is compared to the previous work by \citet{robberto} and is converted into the 2MASS system. 
Completeness corrections are determined and the fractional coverage of the ONC is described. 
\subsection{Source detection and photometry}
Source detection is complicated by strong nebulosity, especially close to the  centre  of the ONC, and by the under--sampled PSF of   the NICMOS Camera 3. 
A ring median filtered image was subtracted from each mosaiced image in order to remove the large scale nebulosity. 
The inner and outer annuli for the ring filter were 7 and 11 pixels, (1\farcs 4 -- 2\farcs 2). 
This makes source detection easier but the photometry was performed on the frames with the nebulosity maintained. 
The PSF is better sampled  in the \filter{F160W} images than the \filter{F110W} images due to the larger full width at half maximum at longer wavelengths. 
Since the \filter{F160W} data are also deeper than the \filter{F110W} data, we have used the \filter{F160W} data  for the source detection. 

Point sources were  visually identified in each individual frame. 
The radial profile of each source  was inspected in order to exclude spurious detections due to nebulosity. 
We detect a total of 2054 objects. 
The positions of the objects detected in the \filter{F160W} images were then used for the photometry in the \filter{F110W} band. 
The photometry was performed with an  aperture radius of 2.5 pixels (0\farcs5)  and a sky annulus between 10 and 15 pixels (2\arcsec-3\arcsec) for both filters. 
Zero points were adopted from the HST calibration web pages.
Bright isolated stars in different regions were used  to determine  aperture corrections. 
The random error on the aperture correction was 0.02 mag for both bands. 
The adopted zero points, including the aperture corrections, are 22.50 mag for the \filter{F110W} band observations and 21.66 mag for the \filter{F160W} band observations. 
Objects  in common with the dataset by \citet{robberto}  were used to calibrate the astrometry of each mosaic. 
For one of the mosaics, no ISPI star was present within the field--of--view. 
Coordinates for the stars in this field were obtained from the header information and is expected to be accurate to $\sim1$\arcsec, the average accuracy of the header coordinates for the mosaics that could be calibrated with ISPI. 
The astrometry is in agreement with the ISPI astrometry to within 0\farcs1 for 60\% of the sources and within 0\farcs2 for almost 90\% of the sources. 
The sources and their photometry are presented in Table~\ref{source_table}.

\begin{table*}
\caption{The sources detected in the  NICMOS Camera 3 mosaics. Objects located in frames that could not be astrometrically calibrated are marked. Magnitudes are in VEGAMAG. The full table is electronically available on A\&As web site.}
\centering
\begin{tabular}{lllllll}
\hline
\hline
ID & RA & DEC & F110W & eF110W & F160W & eF160W \\
     & (J2000) & (J2000) & (mag) & (mag) & (mag) & (mag) \\
   0 &83.55908 & -5.49581 & 23.71 &  0.80 & 21.23 &  0.36\\
    1 &83.55791 & -5.49974 & 22.53 &  0.44 & 19.84 &  0.18\\
    2 &83.55621 & -5.50233 & 19.18 &  0.09 & 17.09 &  0.05\\
    3 &83.55345 & -5.50198 & 23.53 &  0.70 & 21.12 &  0.33\\
    4 &83.55041 & -5.50002 & 23.13 &  0.58 & 20.45 &  0.24\\
    5 &83.54137 & -5.50148 & 22.86 &  0.53 & 20.62 &  0.27\\
    6 &83.54234 & -5.49987 & 23.26 &  0.65 & 20.50 &  0.25\\
    7 &83.54798 & -5.50549 & 14.19 &  0.01 & 13.06 &  0.01\\
    8 &83.54616 & -5.50673 & 22.85 &  0.52 & 21.60 &  0.43\\
    9 &83.54810 & -5.50938 & 20.40 &  0.16 & 18.29 &  0.09\\
\label{source_table}
\end{tabular}
\end{table*}

\subsection{Integrity of the photometry} 
\begin{figure}
\centering
\includegraphics[width=8cm]{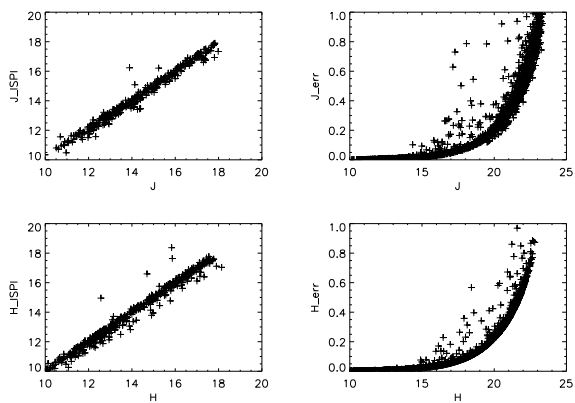}
\caption{Upper and lower left: A comparison of the NICMOS photometry converted into the 2MASS system with the photometry obtained with ISPI, in the 2MASS system \citep{robberto}. 
Only objects with photometric errors less than 0.1 mag in both filters are shown. 
Upper and lower right: The NICMOS photometric errors as a function of magnitude.}
\label{phot_compare}
\end{figure}

The NICMOS photometry has been calibrated into the 2MASS system utilizing the ISPI data which was transformed into the 2MASS system \citep{robberto}. 
We chose relatively well exposed sources, not located close to the edge of the NICMOS mosaics. 
Least square linear fits were performed to the \filter{F160W} band magnitudes and the \filter{F110W}--\filter{F160W} color. 
\begin{eqnarray}
{H_{2MASS}} & =  & \filter{F160W}+a_{0H}\\
{J_{2MASS}} & = & \filter{F110W}+a_{0J}+a_{1J}*(\filter{F110W}-\filter{F160W}),  
\end{eqnarray}
where $a_{0H}=-0.25\pm0.01$, $a_{0J}=-0.26\pm0.01$, and $a_{1J}=-0.35\pm0.01$, respectively.
Fig.~\ref{phot_compare} shows the ISPI photometry as a function of the NICMOS photometry for both the \filter{J} and \filter{H} bands. 
All sources with a nominal photometric error smaller  than 0.1 mag in each band both in this study and the ISPI observations are shown. 
Further, the photometric error for the NICMOS sources is shown for both filters. 
The agreement with the ISPI photometry is in general good but there is some scatter. 
This is to be expected due to the difference in spatial resolution making it easier to separate starlight and nebulosity in our observations but also due to our poorly sampled PSF, especially in the \filter{J} band. 
In any case, our photometry is significantly deeper than the previous ISPI study, which reached 3$\sigma$ limits of \filter{J}$\sim19.5$ mag, and \filter{H}$\sim 18$ mag. 
The faintest sources identified in this study have magnitudes of  \filter{J}$\sim 22.0$ mag, and \filter{H}$\sim 21.0$ mag for the areas with limited nebulosity. 
The discussion below describes in more detail how the completeness limit depends on the distance from the cluster centre, and hence on the amount of nebulosity.

\subsection{Completeness corrections}
We have performed completeness experiments to determine the fraction of stars detected as a function of object magnitude. 
Bright isolated stars from several frames were used to create a model for the point spread function. 
Artificial point sources were then placed randomly within the observed frames, with \filter{F160W} magnitudes randomly chosen  between 15 and 20 mag. 
In each frame around 80\% of the detected number of stars were  added as artificial stars. 
The frames are not crowded and the artificial star experiments are being performed to test the detection limit as a function of nebulosity. 
The relatively large number of stars added per frame does not influence this limit. 
A color of \filter{F110W}-\filter{F160W}=2 mag, was adopted for all the artificial stars, 
corresponding to the colour of the faintest objects within the extinction limited sample defined below (see Fig.~\ref{CMD}).

The artificial stars were recovered by hand in the \filter{F160W}\ frames in a similar manner to the real star detection. 
Photometry was then performed in both the \filter{F160W}\ and the \filter{F110W}\ frames. 
An artificial star was considered detected if the magnitude error in both filters was less than 0.3 mag, similar to the maximum error for some objects used in the extinction limited sample. 

\begin{figure}
\includegraphics[width=8cm]{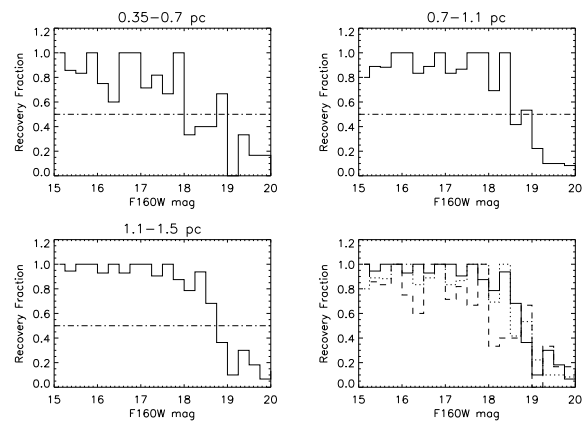}

\caption{The completeness correction for different annuli as a function of \filter{F160W}\ band magnitude. 
From the upper left is shown the completeness corrections for the inner annulus (0.35--0.7pc), the middle annulus (0.7--1.1pc), the outer annulus (1.1--1.5pc), and in the lower--right a comparison of the three annuli.}
\label{completeness_corr}
\end{figure}

Fig.~\ref{completeness_corr} shows the completeness corrections for three different annuli as a function of \filter{F160W}\ band magnitude. 
The 50\% completeness limit clearly varies across the survey area. 
The presence of strong nebulosity close to the cluster centre results in a brighter 50\% completeness limit, as expected. 
In particular, within the central 0.35pc radius, the strong nebulosity and undersampled PSF makes the detection of fainter point sources difficult and the completeness corrections are substantial. 
On the other hand, farther than 0.35pc from the centre the completeness is high down to \filter{F160W}=18 mag. 
For comparison, a 1 Myr 20M$_{jup}$ object at the distance of the ONC has  \filter{H} band magnitude of 16.4 mag. 
Thus, even at 0.35pc we can detect essentially  all  1 Myr objects down to 20M$_{jup}$ seen through an extinction of A$_\mathrm{V}=7$ mag (see below), adopting the models by \citet{baraffe}. 
At larger radial distances objects below 10M$_{jup}$ are detected.  
Due to the incompleteness within 0.35pc we focus our analysis on the objects at larger radii. The current survey thus complements the observations by \citet{HC00} and \citet{luhman_ONC} that both cover the central 0.6pc square region. 
For all radii the completeness limit is higher than the detection limit of the data, which is due to both the nebulosity and to the structure in the nebulosity that can hide in particular faint sources.

\subsection{Fraction of the ONC covered}
\begin{figure}
\includegraphics[width=8cm]{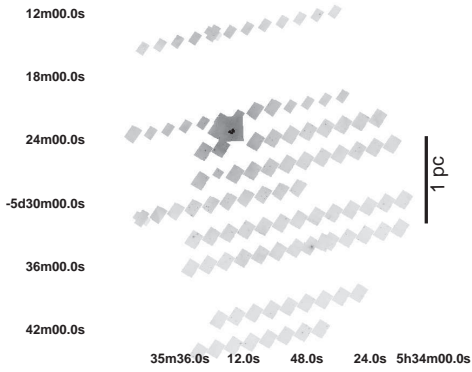}

\caption{\filter{F160W} band  26\arcmin$\times$33\arcmin\ mosaic of the observed region within the ONC. For reference, the central 140\arcsec$\times$140\arcsec\ region observed by \citet{luhman_ONC} is included. 
 The location of the Trapezium stars are marked by small boxes in the mosaic.} 
North is up and East is to the left.  
\label{show_area}
\end{figure}

\begin{figure}
\includegraphics[width=8cm]{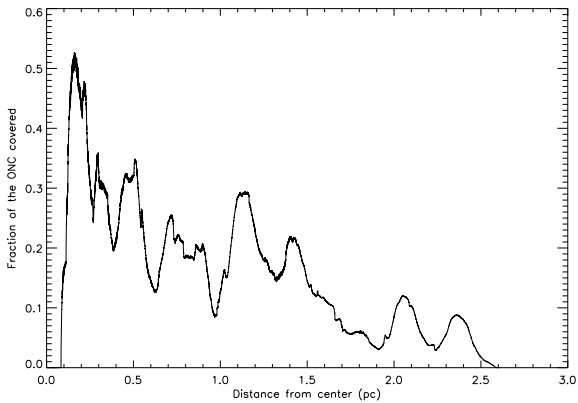}
\caption{The fraction coverage of the ONC by the NICMOS Camera 3 observations. 
The fraction covered is shown as a function of radial distance from the centre ($\theta^1$ Ori C). }
\label{frac_area}
\end{figure}

Since the observations were obtained in parallel, the field--of--view covered is not contiguous. 
The very centre of the cluster is not covered and the majority of the observations are to the south and south--south--west of the cluster centre. 
 Fig.~\ref{show_area} shows the location in the ONC of each individual mosaic. 
For reference, the mosaic observed by \citet{luhman_ONC} is inserted  and the Trapezium stars marked. 
Fig.~\ref{frac_area} shows the fractional area of the ONC covered as a function of distance from the cluster centre. 

 The total area imaged is 162 square arcminutes, corresponding to 2.35 $\mathrm{pc^2}$ for  a cluster distance of 414pc. 
The coverage extends in radius from 0.05pc out to 2.55pc from $\theta^1$ Ori C. 
Due to the large region of sky covered  with a small filling factor, the coverage is relatively low, 20--50\% within 0.5pc, 10-30\% between 0.5pc and  1.5pc, and less at larger distances as shown in Fig~\ref{frac_area}. 

\section{Results and Analysis}
We now present the J--H versus H color--magnitude diagrams for the surveyed region and   estimate the contamination due to field stars. 
The mass of each object is derived from its de--reddened H band magnitude. 
We create an extinction limited sample to ensure an unbiased selection of cluster members as a function of object mass  before determining  the ratio of low--mass stars to brown dwarfs for the ONC as a function of radius.

\subsection{The color--magnitude diagram}
\begin{figure}
\includegraphics[width=8cm]{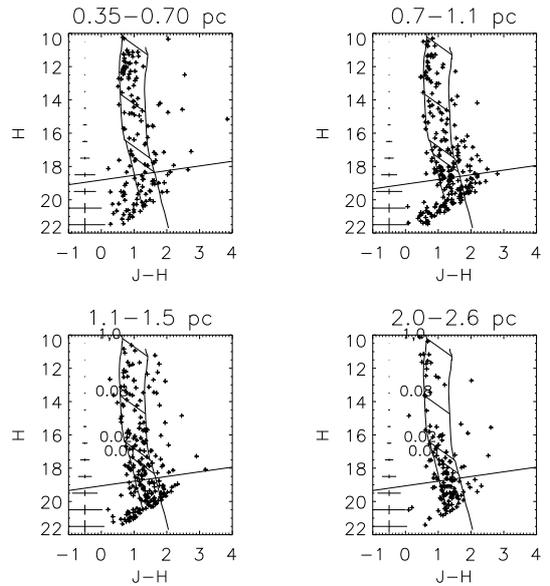}
\caption{The color--magnitude diagram for the surveyed region. Overplotted are a 1 Myr isochrone \citep{baraffe} shifted to the distance of the ONC,  the same isochrone reddened by A$_\mathrm{V}=$7 mag, and the 50\%\ completeness limit for the data. 
Masses for a 1 Myr population of 1.0, 0.08, 0.02, and 0.01M$_\odot$ are indicated on the lower left panel. 
The region used as control field is shown in the lower right corner (2.3--2.6pc). 
The typical photometric and colour error are shown in each diagram as a function of object magnitude.}

\label{CMD}
\end{figure}

Fig.~\ref{CMD} shows the color--magnitude diagram for the surveyed region outside 0.35pc. 
The data have been divided into the annuli that will be used below to derive the ratio of low--mass stars to brown dwarfs. 
Overplotted are the 50\%\ completeness limits, a 1 Myr isochrone from  \cite{baraffe}  converted into the 2MASS system using the same procedure as in \citet{andersen06} with the filter transformations by \citet{carpenter_2mass},  and the same isochrone reddened by A$_\mathrm{V}=7$ mag. 

Most of the objects from 1M$_\odot$ to the detection limit below 0.01M$_\odot$  are located between the 1 Myr isochrone shifted by the distance modulus to the ONC, and the same isochrone reddened by A$_\mathrm{V}=7$ mag. 
The objects on the blue side of the 1 Myr isochrone are likely foreground stars or a significantly older population.

\subsection{Extinction limited samples}
One complication in calculating the IMF in young star forming regions is the presence of differential extinction. 
Objects are reddened by different amounts due to patchy obscuration along the line of sight. 
As evident from Fig.~\ref{CMD}, this is the case for the ONC. 
Higher mass objects can be detected through larger amounts of extinction than lower mass objects due to their intrinsic brightness. 
As illustrated in e.g. \citet{andersen09} this can potentially introduce a spurious deficit in the derived number of low mass objects.
One approach to avoid this bias is to construct an extinction-limited sample, which is obtained by considering only objects within a certain range of extinction while taking the completeness limit into account. 

 The lower mass limit for the extinction-limited sample is set by a combination of depth of the data, completeness corrections, and maximum extinction allowed. 
The choice of an extinction limit is a compromise between completeness and the low mass limit (lower extinction) and maximizing the sample size (higher extinction). 

The extinction for each individual object is determined by shifting the object in the color-magnitude diagram back to the assumed isochrone, in a similar manner as in \citet{andersen06}. 
We find that more than 80\% of the objects are affected by  an extinction of A$_\mathrm{V}=7$ mag or less and use this limit for our extinction limited sample. 

\begin{figure}
\includegraphics[width=8cm]{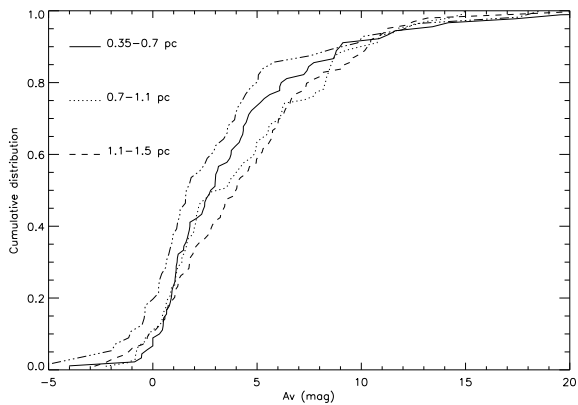}
\caption{The derived extinction for sources in the ONC within the magnitude limits $10.5 < H < 18$ mag, which corresponds to the range from where the ratio of stars to brown dwarfs is determined. }
\label{cumu_ext}
\end{figure}

We have examined whether the extinction distribution is similar at different distances from the centre of the ONC.  
Fig.~\ref{cumu_ext} shows the cumulative distribution of derived extinction for objects in the magnitude range $10.5 < H < 18$ mag, similar to the magnitude range from where the IMF is derived. 
There is no obvious trend of the extinction distribution as a function of radius. A two--sided Kolmogorov--Smirnov test between the cumulative distributions for the different radii indicates that we cannot rule out that the underlying distribution is the same. 
 The maximum deviations between the extinction distributions for different radii were 0.12 with the significance levels varying between 0.37 and 0.62. 
Thus, since the distribution of extinction appears to be similar as a function of radius, the same extinction limit is  used for all radii. 

The mass of each object is then determined by sliding it along the reddening vector in the colour-magnitude diagram back to the isochrone. 
Since only \filter{J} and \filter{H} band observations are available we have no knowledge if an object is associated with a circumstellar disc that would give  rise to an infrared excess. 
With \filter{JHK} photometry the disk is typically taken into account by de--reddening the objects to the T-Tauri locus in the colour--colour diagram \citep{meyer_disk}

Not taking disks explicitly into account will tend to overestimate the derived mass compared to the actual mass. 
Thus, a strong radial dependence on the disk fraction could in principle mimic a decreasing ratio of stars to brown dwarfs as a function of radius. 
\citet{hillenbrand_disk} find  a modest decrease of the disk fraction, from some 65\% showing strong evidence of a disk in the centre to 45\% at a radius of 1.1pc. 
The ratio determined from the central 0.3pc suffers less from the effects of disks. 
\citet{slesnick}  determined the spectral types through spectroscopy and although veiling is still a potential issue, the use of \filter{J} band spectra reduces this effect since the photosphere for late type objects peak in this wavelength range. 
However, the bolometric magnitudes still depend on the de--reddening in the colour--magnitude diagram. 
Using different de--reddening approaches provide difference in the luminosity of $\Delta log L=0.05$ and is not expected to have an influence on the ratio of stars to brown dwarfs. 
The colour--magnitude diagrams in Fig.~\ref{CMD} further indicate this is not likely to be the source of the decrease in the ratio of stars to brown dwarfs as a function of radius. 
We would expect to see the strongest effect is in the 0.35--0.7pc bin where few objects are located close to the star--brown dwarf division line. 
For objects at the star--brown dwarf boundary the excess is  \filter{J-H} of 0.1 mag or less \citep{meyer_disk},   which after de-reddening will correspond to  maximum shift in  H band magnitude of 0.16 mag. 
At most 4 objects would be redefined from stars to brown dwarfs from 0.35-0.7pc assuming all objects had disks and all objects were affected by 0.16 mag of excess.  
Several objects at larger distances could also potentially be re-classified, 1 object from 0.7-1.1pc, and 3 objects from 1.1-1.5pc.
Neither will change the ratios significantly and according to \citet{hillenbrand_disk} half or less of the objects would show evidence for an infrared excess. 
We thus conclude it is unlikely the decrease in the ratio of stars to brown dwarfs is due to a lower disk fraction at larger radii. 

\subsection{Field star contamination}
Due to the relatively large area covered we expect some field star contamination. 
The contamination can be separated into two types, namely  that of foreground and background stars. 
Since the ONC is located at only 414pc, we expect the foreground field star contamination to be relatively minor. 
In addition, the foreground objects will have relatively blue colors compared to the cluster members (see below). 

Background stars are a larger concern. 
Two effects limit the contamination. 
First, the ONC is located at $(l,b)=(209.07,-19.49)$, out of the Galactic plane and away from the inner Galaxy. 
 Further, the ONC and surrounding region is associated with and partly in front of a giant molecular cloud complex acting  as a shield for background objects. 
By comparison with the $^{13}CO$ line map from \citet{bally87} it is evident that all the observed region is located within the $\int$ shaped filament in Orion A--North, in agreement with strong emission seen in the IRAS 100 $\mu m$ images. 
Based on the strength of the CO emission the extinction averaged over a 40\arcsec beam is well above A$_\mathrm{V}=15$ mag, in agreement with the low number of reddened stars seen even far from the centre of the ONC. 

We have estimated the field contamination in  the following way. 
Since the cluster surface density at larger radii decreases as $r^{-2}$ \citep{HH}, we expect the outer region  (2.0-2.6pc) to have a relatively small number of cluster members. 
Thus, the outer regions can be used as a conservative over--estimate of the field star contamination. 
However, as we indicate below there appears to be an increase in the relative number of brown dwarfs at larger radii. 
Any cluster population present at large radii will therefore result in an over subtraction of the brown dwarf content. 
The derived ratio of stars to brown dwarfs after correction based on the outer regions will thus be an upper limit. 

 A Galactic model could in principle be used to predict the field star contamination along the line of sight of the ONC \citep[e.g.][]{robin}. 
However, in practice this is difficult due to the lack of calibration of the models for faint magnitudes. 
The model can be used as a test of whether the control field reproduces the expected number of foreground field stars. 
Using the Galactic model by \citet{robin} we find the predicted number of foreground field objects towards the control area (2.0--2.6pc) to be 8.9. 
The measured number of blue objects (\filter{J}--\filter{H} $<$0.8 mag) within the magnitude range $10 < H < 19$ mag is 10, suggesting that the outer annulus traces the field star population well. 
Based on the agreement with the foreground population and the similarity in the extinction distribution as a function of radius, we adopt the outer annulus, 2.0-2.6pc as the control region for both foreground and background stars. 

\subsection{The low--mass Initial Mass Function}
From the extinction limited samples we can construct an unbiased estimate of the IMF down to the completeness limit of the data. 
As illustrated in  Fig.~\ref{completeness_corr} the completeness in our inner bin decreases sharply below  \filter{H}=18mag but individual objects are still detected down to \filter{H}=20mag. 
Thus, by restricting the sample to an extinction limit of A$_\mathrm{V}=7$ mag and a lower mass limit of 20M$_\mathrm{Jup}$ the sample is expected to be complete across the whole mass range for all radii. 
The completeness tests showed the inner annulus is incomplete for the 10--20M$_\mathrm{Jup}$ mass range and this part of the mass function is thus an upper  limit. 
In addition, as described below, the field contamination is substantial for the lowest mass range. 
We thus mainly focus the discussion of the IMF down to 30M$_\mathrm{Jup}$ but present the corresponding ratios for the samples extending to 10 and 20M$_\mathrm{Jup}$.

\begin{figure}
\includegraphics[width=8cm]{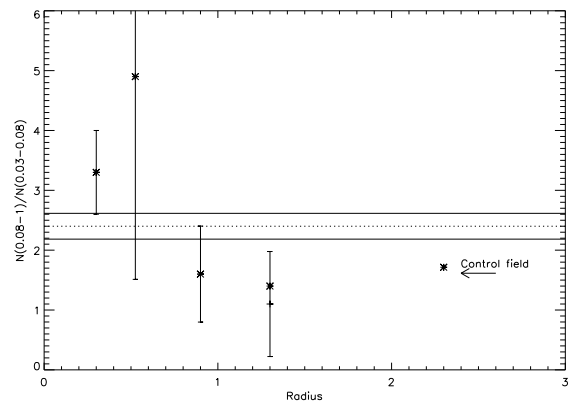}

\caption{The ratio of low--mass stars (0.08--1.0M$_\odot$) to brown dwarfs (0.03--0.08M$_\odot$) for the ONC as a function of radius. 
The plus signs with error bars indicate the field star corrected measurements whereas the squares indicate the measurements before field star subtraction. 
The inner data point is from \citet{slesnick} and the horizontal line is the average ratio with the 1$\sigma$ error interval provided by the dashed lines.}
\label{ratio_radius}
\end{figure}

Following \citet{meyer00} and \citet{andersen08} we have determined the ratio of low--mass stars (0.08--1M$_\odot$) to brown dwarfs both for the cluster as a whole, and for 3 separate annuli outside a 0.35 pc radius. 
The results are summarized in Table~\ref{ratio_summary} and the ratio of stars to brown dwarfs as a function of radius is shown in Fig.~\ref{ratio_radius} where the ratios both with and without field stars subtracted are shown. 

The ratio determined at 0.35--0.7 pc is consistent with the results by \citet{slesnick} for the central 0.3 pc radius of the cluster. 
However, for larger distances there is a drop in the ratio of stars to brown dwarfs. 
Outside a radius 0.7 pc the ratio drops by almost a factor of two, although the error bars for the individual annuli are relatively large. 
The total ratio of stars to brown dwarfs for the cluster as a whole has been calculated by correcting the source counts in the different annuli by the fraction of the cluster covered by that annulus (derived from Fig.~\ref{frac_area}). 
As shown by \citet{HH}  the surface density is well represented by a King profile, and thus extends as $r^{-2}$ outside the core radius. 
Thus, even though the surface density is lower at larger radii, this is compensated by an increase in the area in each annulus. 
The low ratio far from the centre drives the global ratio to a relatively low value of $R_{03}=2.4\pm0.2$, lower than  determined in the cluster centre. 

The drop in the ratio of stars to brown dwarfs as the distance from the centre increases is also seen for the sample with a mass limit of 20M$_\mathrm{Jup}$ or 10M$_{Jup}$  instead of 30M$_\mathrm{Jup}$ but due to the stronger contamination of field stars as the sample is extended to lower masses we focus the analysis on the higher mass limit data. Whereas the field contamination is some 10\%\ for the brown dwarf content  down to a mass limit of 30M$_\mathrm{Jup}$, it increases at lower masses and hence fainter magnitudes. 
This is a consequence of the underlying luminosity function for the field star population  which extends as a power--law to faint magnitudes.

 Could the lower ratio of stars to brown dwarfs as a function of distance from the ONC centre be due to age effects? 
Adopting the new lower distance to the ONC \citep{menten} and a more extensive sample of members where the spectral type has been determined through spectroscopy \citep{hillenbrand97} or through narrow band imaging of TiO absorption features, \citet{dario} suggest that the cluster might be older than the previous estimate of 1 Myr.  Recent work has further suggested an age gradient across the ONC \citep{reggiani}, where the centre is dominated by a younger population whereas a wider age distribution is seen at larger radii with older stars having a larger average distance from the centre. 

The effect of an older cluster at the same distance is that the pre--main sequence stars and brown dwarfs gets dimmer, an effect that is stronger for lower masses and it could thus replicate a decrease of the ratio of stars to brown dwarfs if the age adopted for the outer parts of the cluster is too low. 
\citet{reggiani} find an average age for the (stellar) population between 1 pc and 1.5 pc to be $\log (age=6.3)$. 
We have therefore also calculated the ratios of stars to brown dwarfs in the outer annuli for an age of 2 Myr. 
We find the ratio of low--mass stars  to brown dwarfs only changes slightly from the values determined for a 1 Myr isochrone, $R_{03}(2Myr, d=1.1-1.5pc))=1.1\pm08.$, and R$_{02}(2Myr,d=1.1-1.5pc)=0.8\pm0.6$ and this will thus not change the conclusion that the ratio of stars to brown dwarfs decrease as a function of radius.

\section{Discussion}
We present the constraints that our observations place on the global IMF for the ONC and compare the results with the field IMF and results from other nearby star forming regions. 
Evidence for variations in the IMF as a function of environment is discussed. 
\subsection{Variations in the ratio of stars to brown dwarfs as a function of radius}
The decreasing ratio of stars to brown dwarfs as a function of radius suggests that the cluster is mass segregated. 
A similar conclusion was found by \citet{HH} for the massive stars in the cluster centre. However, there was no indication for mass segregation for the lower mass population as their study only extended to 0.3M$_\odot$ and thus compared high mass stars to low-mass stars. 
In contrast, we find the sub--stellar  members to be segregated compared to the stars as a whole, which are dominated by the lower mass members. 

\citet{HH} concluded that it is possible for the high-mass  segregation to  have an dynamical origin. 
This is not the case for the brown dwarfs observed here with the current cluster surface density profile. 
The half mass relaxation time was estimated in \citet{HH} to be 6.5 Myr which is much longer than the age of the cluster (1-3 Myr) and thus unlikely to cause the variations in the ratio of stars to brown dwarfs through dynamical 2 body  evolution \citep[c.f.][]{bonnell98}.

A decrease of the ratio of stars to brown dwarfs as a function of radius has been observed in some open clusters as well, e.g. IC 2391 \citep{bourdreault} where the brown dwarfs were found to have a flatter distribution than the low--mass stars. 
Similarly, for IC 348 and central parts of the ONC, \citet{kumarschmeja} use the minimum spanning tree to suggest the brown dwarfs are less centrally concentrated than the stars. 
A flatter distribution for brown dwarfs compared to stars is what would be expected from an ejection scenario where a fraction of the brown dwarfs are due to ejections in multiple systems during the earliest phases of star formation \citep[e.g.][]{reipurthclarke,bate03}. 

 More recent simulation by e.g. \cite{bate09,bonnell10} does not predict an overabundance of brown dwarfs further away from the centre when the simulations end. 
However, the simulations typically end after 1--1.5 free fall times which, depending on the specific simulation corresponds to  0.2-0.5 Myr, significantly younger than the current ONC. 
Thus, a more extended brown dwarf distribution could be generated by subsequent dynamical interactions in the core leading to a substantial unbound population or a higher overall velocity dispersion.

\subsection{Comparison with other Star forming regions}

 A smaller ratio of the number of stars to brown dwarfs for the whole cluster compared to the centre region has several implications. 
\citet{bate09} predicted a value of $R_{03}=1.25$, based on SPH simulations of a spherical collapsing cloud producing a total of 1254 objects, even lower than we determine here for the ONC. 
 It is worth noticing that our value of $R_{03}=2.4\pm0.2$ is not consistent with a Chabrier IMF, which predicts too few brown dwarfs  ($R_{03}(Chabrier)=4.95$). 
The probability of the ratio to be drawn from an underlying population with a Chabrier distribution is less than 1\%. 
The ratio we determine is in better agreement with a flat or  rising  IMF  in linear units into the brown dwarf regime. 
We stress however, that our IMF is determined from a photometric sample only and that spectroscopic confirmation of the brown dwarf candidates is highly desirable. 

\citet{andersen08} presented a compilation of well determined ratios of stars to brown dwarfs in nearby young star forming regions and the Pleiades. 
The measurements did not show evidence for a variable underlying IMF as a function of star formation environment. 
Results for some OB associations, e.g. Upper Sco, and $\sigma$ Ori,  have indicated that the derived mass function is richer in brown dwarfs compared to the young cluster sample \citep[e.g.][]{lodieu,caballero}. 
The derived low--mass slope is only weakly falling or flat into the brown dwarfs regime (a slope of $-0.4\pm0.2$ in logarithmic units). 
For both the upper Sco survey and for $\sigma$ Ori, the mass function was determined over a large field of view, several square degrees and the associations are thus probed to much larger radii than e.g. previous studies of the ONC.
However, although for some of the embedded clusters only the central regions have been observed (Mon R2, ONC, NGC 2024), others have much wider regions surveyed and the ratio of stars to brown dwarfs is still relatively high (e.g. Taurus, IC 348).

The current finding of a possible change in the ratio of stars to brown dwarfs at larger radii  suggests  further studies of other nearby regions to larger radii than previously surveyed are needed. 
Also, obtaining kinematic data  for the individual low--mass members to better characterise any mass segregation would be very valuable.  
Current near--infrared large field of view capabilities both from the ground and space are well suited for these studies.

\section{Conclusions}
We have presented deep NICMOS III \filter{F110W} and  
 \filter{F160W} non--contiguous observations of the Orion Nebula Cluster region. 
The data reach  \filter{J}=$22$ mag, and \filter{H}$=21$ mag, deeper than previous large scale high spatial resolution surveys. 
From the data we derive the ratio of low--mass stars (0.08-1M$_\odot$) to brown dwarfs as a function of radial distance from the centre. 
The ratio of stars to brown dwarfs is lower than in the centre, falling from a value of 3.7 in the centre to 1.1$\pm$0.8 in the annulus 1.1--1.5 pc. 
This shows the low--mass content of the cluster is mass segregated to large radii. 
The age of the cluster is much shorter than the half mass relaxation time suggesting the mass segregation for the low--mass population is primordial. 
The number of stars to brown dwarfs for the cluster as a whole is found to be 
$\frac{N(0.08-1M_\odot)}{N(0.03-0.08M_\odot)}=2.4\pm0.2$, which is in better agreement with a flat or rising low--mass IMF than a Chabrier IMF.

\begin{table*}
\caption{The ratio of stars to brown dwarfs in an extinction limited sample with A$_\mathrm{V}=7 mag$ as a function of lower mass limit. Both the ratio corrected for field stars and uncorrected is provided. 
For the inner 0.35 pc, the ratio of stars to brown dwarfs is taken from \citet{slesnick}. The \citet{slesnick} study only extends to 20 M$_\mathrm{Jup}$ and the total ratio of stars to brown dwarfs down to 10 M$_\mathrm{Jup}$ is incomplete for the inner 0.35 pc. }
\begin{tabular}{cccccccccccc}
\hline
\hline
& &  Corrected for field stars & & & Uncorrected & \\ Ratio  & 0.35-0.7pc & 0.7-1.1pc &1.1-1.5pc& Cluster & 0.35-0.7pc & 0.7-1.1pc &1.1-1.5pc & Field\\
$\frac{N(0.08-1)}{N(0.03-0.08)}$ &7.2$\pm$5.6&1.6$\pm$0.8&1.1$\pm$0.8&2.4$\pm$0.2& $\frac{          44}{           9}=$4.9$\pm$1.8&$\frac{          31}{          19}=$1.6$\pm$0.5&$\frac{          27}{          19}=$1.4$\pm$0.4& $\frac{12}{7}$ \\
$\frac{N(0.08-1)}{N(0.02-0.08)}$ &4.4$\pm$2.8&1.4$\pm$0.7&0.8$\pm$0.6&1.7$\pm$0.1&$\frac{          44}{          15}=$2.9$\pm$0.9&$\frac{          31}{          25}=$1.2$\pm$0.3&$\frac{          27}{          29}=$0.9$\pm$0.2& $\frac{12}{12}$ \\
$\frac{N(0.08-1)}{N(0.01-0.08)}$ &4.1$\pm$3.3&1.4$\pm$0.8&1.2$\pm$1.2&1.6$\pm$0.1&$\frac{          44}{          20}=$2.2$\pm$0.6&$\frac{          31}{          32}=$1.0$\pm$0.3&$\frac{          27}{          36}=$0.8$\pm$0.2& $\frac{12}{20}$ \\
\label{ratio_summary}
\end{tabular}
\end{table*}

\end{document}